%% file: main.tex
\PassOptionsToPackage{dvipsnames}{xcolor}

\documentclass[sigconf, screen]{acmart}

\usepackage{xspace}

\usepackage{listings}

\AddToHook{env/lstlisting/begin}{\fboxsep=0pt}

\usepackage{listings-rust}

\usepackage{enumitem}

\usepackage{pdfpages}

\usepackage{commands}

\usepackage{multirow}

\usepackage{cleveref}

\usepackage{colortbl}

\usepackage{microtype}

\usepackage{subcaption}
\DeclareCaptionLabelFormat{mydefault}{(#2)}
\DeclareCaptionLabelFormat{rewriterule}{Rule~#2}
\DeclareCaptionLabelFormat{table}{Table~#2}

\usepackage{code-styles}

\usepackage{tcolorbox}
\NewDocumentCommand{\calloutbox}{O{}m+m}{%
  \begin{tcolorbox}[%
    rounded corners,
    colframe=gray!50, 
    boxsep = 0pt,
    left = 7pt,
    right = 7pt,
    top = 5pt,
    bottom = 5pt,
    colback=gray!10, 
    box align=center,
    valign=center,
    title = {},#1]
    #3%
    \end{tcolorbox}%
}
\AtBeginDocument{%
  \providecommand\BibTeX{{%
          Bib\TeX}}}

\setcopyright{cc}
\setcctype{by-nc-nd}
\acmDOI{10.1145/3832783.3837412}
\acmYear{2026}
\copyrightyear{2026}
\acmISBN{979-8-4007-2882-2/2026/10}
\acmConference[ASE '26]{Proceedings of the 41st IEEE/ACM International Conference on Automated Software Engineering}{October 12--16, 2026}{Munich, Germany}
\acmBooktitle{Proceedings of the 41st IEEE/ACM International Conference on Automated Software Engineering (ASE '26), October 12--16, 2026, Munich, Germany}
\acmSubmissionID{ase26main-p5-p}
\received{2026-03-09}
\received[accepted]{2026-06-18}

\def\BibTeX{{\rm B\kern-.05em{\sc i\kern-.025em b}\kern-.08em
    T\kern-.1667em\lower.7ex\hbox{E}\kern-.125emX}}
\begin{document}

\title{Translation Tag Team: Formal Rules and LLMs Translate More Macros Together Than Apart}

\author{Brent Pappas}
\correspondingauthor
\orcid{0009-0003-0780-743X}
\affiliation{%
  \institution{University of Central Florida}
  \city{Orlando}
  \country{USA}
}
\email{brent.pappas@ucf.edu}

\author{Joseph Zalusky}
\orcid{0009-0000-4065-0380}
\affiliation{%
  \institution{University of Central Florida}
  \city{Orlando}
  \country{USA}
}
\email{josephzalusky@ucf.edu}

\author{Zachary Burkett}
\orcid{0009-0003-7776-3655}
\affiliation{%
  \institution{University of Central Florida}
  \city{Orlando}
  \country{USA}
}
\email{zachary.burkett@ucf.edu}

\author{Paul Gazzillo}
\orcid{0000-0003-1425-8873}
\affiliation{%
  \institution{University of Central Florida}
  \city{Orlando}
  \country{USA}
}
\email{paul.gazzillo@ucf.edu}


\begin{abstract}
	\input{abstract.tex}
\end{abstract}



\maketitle

\section{Introduction}\label{sec:introduction}
\input{introduction.tex}

\section{Translating Macros with \tool{}}\label{sec:translating-macros}
\input{translating.tex}

\section{\tool{} Implementation}\label{sec:implementation}
\input{implementation.tex}
\section{Macro Translation Benchmark}\label{sec:benchmark}

\input{benchmark.tex}

\section{Evaluation}\label{sec:evaluation}
\input{evaluation.tex}

\section{Threats to Validity}\label{sec:threats}
\input{threats.tex}
\section{Related Work}\label{sec:related-work}
\input{related-work.tex}


\begin{acks}
    \input{acknowledgments.tex}
\end{acks}

\section*{Data-Availability Statement}\label{sec:data-availability}
\input{data-availability.tex}

\balance{}
\bibliographystyle{ACM-Reference-Format}
\bibliography{refs.bib}

\end{document}

%% file: abstract.tex
Modern critical software infrastructure is largely written in C.
Since C lacks memory safety, researchers are investigating automatic translation of C to safer languages like Rust.
%
%
But real-world C software consists of more than just C code, often using named code fragments called macros which are not part of the C language proper.
State-of-the-art techniques avoid translating macros by preprocessing C code first before translating it.
But this approach produces translations that are dissimilar to the original C code, because preprocessing inlines all macro definitions.
%
%
%
To preserve macro usage in translated code, we study the language features that macros and C share and distill them into the first formally-specified translator, \tool{}.
%
%
To evaluate \tool{}, we introduce the first macro translation benchmark, \benchmark{}, with test cases based on macros randomly sampled from real-world C programs.
%
We find that \tool{} supports 50\% of \benchmark{}'s macro test cases.
We also use \benchmark{} to evaluate how effective large language models (LLMs)
are at performing the previously-unstudied task of macro translation.
%
%
LLMs translate 22\% to 77\% more of \benchmark{} than \tool{}, but with 8\% and
28\% of these translations being incorrect translations requiring additional
validation by developers.
In contrast, \tool{} only produces correct translations.
%
%
Our key insight is that running \tool{} first then using LLMs on the remainder reaps greater benefits than using either technique alone.
%
This \emph{tag team} approach has an average failure rate \hybridaveragelowerfailureratethanllms{} lower than that of LLMs, while also translating an average of \hybridaveragegreatertranslationsthantool{} more test cases than \tool{}.

%% file: introduction.tex
Modern critical software infrastructure is largely written in C.
%
Since C lacks memory
safety~\cite{back-to-building-blocks,darpatractor,cisa-memory-safety},
researchers are investigating automatic translation of C to safer languages like
Rust~\cite{corrode,c2rust,c2go,3c,c-to-safer-rust,type-migrating-c-to-rust}.
%
However, real-world C code heavily uses~\cite{ernst-et-al, maki} \emph{macros}~\cite{gnu-c-preprocessor-macros}, which are named code fragments and not part of the C language proper~\cite{superc,typechef,cscout,yacfe}.
Instead, the C preprocessor replaces macros in source code with their defined
code fragments to produce pure C code.
%
%
%
%
C tools have long-struggled with macros~\cite{superc,typechef,cscout,yacfe}, and translators are no exception~\cite{c2rust-issue-16,3c-issue-400,3c-issue-40,3c-issue-439}.
State-of-the-art C translators avoid macros by preprocessing first, then translating the resulting pure C code~\cite{c2rust-faq,shiraishi2024contextawarecodesegmentationctorust,3c-issue-400}.
%
But this approach emits translations which inline all macros and lose their abstractions.

The problem is that by expanding macros first, translators end up translating a version of C code different from the original, and emit translated code that lacks the original code's macro abstractions.
For example, the code below is from the Linux kernel source\footnote{\texttt{linux-v6.2-rc2/drivers/gpu/drm/i915/gt/intel\_gtt.h}}:
\begin{lstlisting}[style=style-c-no-numbered, escapechar=^,xleftmargin=.25in]
gen6_pte_t pte = ^\highlightcode{GEN6\_GTT\_ADDR\_ENCODE(addr)}^ | ^\highlightcodegreen{GEN6\_PTE\_VALID}^;
\end{lstlisting}
Syntactically, \texttt{GEN6\_GTT\_ADDR\_ENCODE(addr)} resembles a C function call and \texttt{GEN6\_PTE\_VALID} a C variable, but both are actually macros.
After preprocessing, the macros are inlined, producing the following C code:
\begin{lstlisting}[style=style-c-no-numbered, escapechar=^,xleftmargin=.25in]
gen6_pte_t pte = ^\highlightcode{((addr) | (((addr) >> 28) \& 0xff0))}^ |
    ^\highlightcodegreen{((u32)(((((1UL))) << (0)) + ((int)(sizeof(struct \{}^
      ^\highlightcodegreen{int : (-!!((sizeof(int) ==}^      
        ^\highlightcodegreen{sizeof(*(8 ? ((void *)((long)(0) * 0l)) :}^        
          ^\highlightcodegreen{(int *)8))) \&\& ((0) < 0 || (0) > 31)));\})))));}^
\end{lstlisting}
%
This barely-recognizable C code, which lacks the original programs's macros, is
what translators ultimately end up translating to a new language, because they
preprocess macros first.
Macros are very common in real-world source code: Linux has over 50 thousand
macro definitions and 500 thousand macro invocations, and C programs average one
invocation every four lines of code~\cite{ernst-et-al}.

There is little prior research on preprocessor macro translation.
%
%
Mennie and Clarke describe a translation technique~\cite{mennie-and-clarke} for
a restricted set of macros that act like C variables, omitting commonly-used
function-like macros, but lack a rigorous evaluation or benchmark, and their
translator's code is publicly unavailable.
Visual Studio supports translating macros to C++ constant expression
declarations~\cite{vs-2017-automatic-macro-refactoring}, but lacking semantic
analysis~\cite{maki} gets translations
wrong~\cite{vs-2017-automatic-macro-refactoring-unaddressed-bug} and is
designed for C++, not C.
LLMs are potentially viable for macro
translation~\cite{gpt-4o,claude-3.5-sonnet,ai-for-code-conversion,shiraishi2024contextawarecodesegmentationctorust,lost-in-translation}
and some are specifically designed for code
translation~\cite{claude-3.5-sonnet-excels-at-coding,o1-preview}.
But to the best of our knowledge, there is no prior study of their ability to
translate macros.
Moreover, there are no published macro translation benchmarks that would enable
evaluating different translators.

We introduce the first formally-specified, semantics-aware macro translator.
%
The key challenge is that while macro usage appears syntactically like C
functions, their semantics can differ greatly~\cite{maki}; translators cannot
in general treat them like C functions and expect to produce equivalent code.
The key insight to our approach is that by identifying those macro usages that
behave like C functions, we can transform such macros to true C functions.
The benefit is that existing C translators can then translate them like any
other C function without special support for macros.
Moreover, downstream C translators can convert C-translated macros into
compile-time abstractions, e.g., Rust inline functions~\cite{rust-inline} or
even Rust macros~\cite{rust-macros}, to translate macros in a way that does not
sacrifice performance.
%
We study macro and C language features and identify the specific conditions
under which macros match the behavior of C functions, variables, or enums.
We define seven formal translation rules that capture these conditions and
implement them in a new tool called \tool{}.

With no published benchmark designed for evaluating macro translation, we
introduce a new benchmark suite called \benchmark{} sampled randomly from
real-world C programs, independent from our formal translation rule
specifications.
%
\benchmark{} is a statistically significant (95\% confidence level, 5\% margin
of error) random sample of \numtestcases{} macro test cases stratified over
\numbenchmarkprograms{} real-world C programs used in prior studies of C and
macros~\cite{ernst-et-al,maki,superc,typechef}.
%
%
%
Each test case focuses on the translation of a single macro and contains its
surrounding language context.
%
\benchmark{} also contains more than one thousand real translations of its test
cases produced during the evaluation in this paper.
%
We semi-automatically checked each translation to record the translation's
correctness, what language construct the tool attempts to translate each macro
to, why each tool failed for incorrect translations, timings, and other
metadata.
We then use \benchmark{} to evaluate how effective popular large language
models (LLMs)~\cite{gpt-4o, claude-3.5-sonnet, o1-preview} are at translating
macros, and compare them to \tool{}.
We specifically compare \tool{} to LLMs, and no other rule-based tools, because
to the best of our knowledge \tool{} is the first publicly-available
semantics-aware macro-to-C translator.
Other rule-based tools either ignore macro semantics and emit incorrect
translations~\cite{vs-2017-automatic-macro-refactoring}, lack a
publicly-available implementation~\cite{mennie-and-clarke}, or do not translate
macros to C~\cite{demacrofication, astec}.
\tool{} translates all \benchmark{}'s test cases in 37 seconds and produces no
incorrect translations, but only emits translations for 50\% of \benchmark{}'s
test cases. Meanwhile, LLMs emit translations for 22\% to 77\% more test cases
than \tool{}, but spend 45 minutes to 2.5 hours on translation, and get 8\% to
28\% of translations wrong.
The presence of incorrect translations means that developers validate all LLM
translations to find correct ones, a substantial task for large codebases like
the Linux kernel which contains thousands of macros.
%
In contrast, \tool{}'s semantics-aware translation rules only produce correct
translations, obviating substantial validation.


We show that using \tool{} to translate macros first, and then ``tagging in'' LLMs to translate the remaining macros, reaps greater benefits than using either technique alone.
We observe an 18\% to 45\% lower failure rate for this \emph{tag team} approach than using
LLMs alone and 30\% to 80\% more translations than using \tool{} alone.
Counterintuitively, although LLMs have a much higher average failure rate on
the remaining macros not supported by \tool{} (\llmaveragefailureratenonmerc{} vs. \llmaveragefailurerate{}), it is
less costly in both overall translation time and post-translation validation
effort to reserve LLMs for these harder cases, rather than treating LLMs as
all-purpose translators.
%



We make the following contributions: \begin{itemize}[noitemsep, topsep=4pt]



	\item A comparison of macro and C semantics and a set of formal translation rules for common macro usage (\Cref{sec:translating-macros}).


	\item \tool{}, an implementation our translation rules (\Cref{sec:implementation}).

    \item \benchmark{}, the first benchmark specifically focused on preprocessor macro translation (\Cref{sec:benchmark}).


	\item An evaluation and comparison of the macro translation capabilities of \tool{} and several popular LLMs (\Cref{sec:evaluation}).

\end{itemize}

The anonymized, publicly-available artifact\footnote{\url{https://doi.org/10.5281/zenodo.21798007}\label{fn:artifact}} includes \tool{}, \benchmark{}, experimental scripts, and data. 

%% file: translating.tex

We present formal translation rules specifying how and when to convert macro definitions into C definitions, and give examples of how \tool{} follows these rules to translate macros.
We designed \tool{}'s translation rules by studying the C preprocessor's ISO
specifications~\cite{c90-standard, c23-standard}, its usage in large open source
projects, and analyses of it in prior work~\cite{cppsig, ernst-et-al, maki}.

%

\subsection{Macro Definitions}


%
%
Each macro definition has a C identifier as a name, and a possibly empty sequence of C tokens as a body.
\textit{Function-like} macro definitions also have a parenthesized, comma-separated list of C identifiers as arguments. \textit{Object-like} macros do not.
For example:

\begin{lstlisting}[style=style-c-numbered] 
#define PI          3.14|\label{ln:olm-flm:olm-def}|
#define ADD(a, b)   a + b|\label{ln:olm-flm:flm-def}|
PI * (ADD(PI, 2)|)\label{ln:olm-flm:calls}
\end{lstlisting}

\noindent{}Lines~\ref{ln:olm-flm:olm-def} and~\ref{ln:olm-flm:flm-def} define the object-like macro \texttt{PI} and the function-like macro \texttt{ADD()}.
Line~\ref{ln:olm-flm:calls} preprocesses to \texttt{3.14 * (3.14 + 2)}.




A macro invocation need not expand to a complete C statement:

\begin{lstlisting}[style=style-c-numbered, caption={\texttt{MY\_IF()} macro definition and invocation.}, label={lst:my-if}]
#define MY_IF(X)    if (X)
MY_IF(x > 0) { ... }|\label{ln:my-if:my-if-call}|
\end{lstlisting}

\noindent{}Line~\ref{ln:my-if:my-if-call} preprocesses to \texttt{if (x > 0)}, a fragment of a
C if statement.
%

\subsection{Determining Translatability}

\input{figures/functions.tex}

\input{figures/venn-diagrams.tex}

\input{figures/translation-rules.tex}

Figure~\ref{fig:venn} presents a six-way Venn diagram comparing the C and C
preprocessor semantic features \tool{} uses to decide which macro definitions
are translatable to C variable, enum, and function definitions, and which macro
arguments are translatable to C function arguments.
A macro that only uses the semantic features of a C construct can be safely
translated to a definition of that construct, without needing to change the
macro's callsites.
For instance, a macro that only uses features in Figure~\ref{fig:venn}'s
overlapping regions between macros and enums would be safe to turn into a C
enum by swapping its macro definition with an enum definition, without changing
its callsites.
%
This enables downstream C translators to translate C code while preserving
macro abstractions, and helps remove needless macros from programs like the
Linux kernel that have guidelines to prefer functions over macros when
possible~\cite{linux-coding-style-macros}.

Figure~\ref{fig:venn} is formalized in Figure~\ref{fig:translation-rules} as a set of inference rules for translating macros to C code.
Each rule has the form:

\[
	\frac{p_0 \cdots p_n}{V_0 \cdots V_n \vdash A \leadsto B}
\]

\noindent{}Where $p_0 \cdots p_n$ are premises that must all be true for the
rule to apply, $V_0 \cdots V_n$ are context variables necessary for evaluating
the rule's premises, and $A$ and $B$ are source and target syntax.
%

%

Rules may use any of the four context variables M, $\Gamma$, A, and $\Delta$.
M and $\Gamma$ map each macro's name to its set of invocations
and the types of the C expressions it expands to, respectively.
A and $\Delta$ map each pair of macro name and argument name
to the set of the argument's expansions and the types of the C expressions it
expands to, respectively.
A macro or macro argument that expands to a non-expression C statement maps to
\textbf{void} in $\Gamma$ or $\Delta$, respectively.

Rule premises use helper functions typeset in \textsc{SmallCaps} font.
%
%
Table~\ref{tab:functions} defines each function informally.
All premises and functions expect named macros expanding to whole C statements,
and \tool{} does not translate macros like \texttt{MY\_IF()}
(Listing~\ref{lst:my-if}) that expand to C syntax fragments, or variadic
macros.
To compute these functions in practice, \tool{} uses an existing macro analyzer
named Maki~\cite{maki}, which works by first hooking into the Clang
preprocessor and abstract syntax tree (AST) to determine which macro
invocations align with AST nodes.
Maki then performs semantic analysis on the AST-aligned invocations to compute
for each of them the functions shown in Table~\ref{tab:functions} as a set of
Boolean properties.
%

\subsection{Translating Object-like Macros}\label{sec:macro-to-var}

Rules~\subref{rule:var} and~\subref{rule:enum} (\Cref{fig:translation-rules}) translate to variables and enums, respectively, macros whose semantics match the rules' premises.
For example, the object-like macro \texttt{QUIT}, which expands to the character \texttt{`q'}, satisfies Rule~\subref{rule:var}'s conditions for translation to a C variable:

\noindent
\begin{minipage}{0.9\linewidth}
\begin{lstlisting}[style=style-c-numbered, caption={\texttt{QUIT} macro
definition and invocation.}, label={lst:quit}]
#define QUIT 'q'|\label{ln:quit-def}|
void check_quit(void) {
    if (QUIT == getch()) exit(0);|\label{ln:quit-call}|
}
\end{lstlisting}
\end{minipage}

Premise~\ref{pr:var:global} checks that \texttt{QUIT} is defined outside of function boundaries, so that translating it to a C variable will not limit it to a specific scope.
Premise~\ref{pr:var:mu} collects all \texttt{QUIT}'s invocations, in this case the one invocation on Line~\ref{ln:quit-call}, into a set $\mu$.
Premises~\ref{pr:var:env} and~\ref{pr:var:meta} check that \texttt{QUIT} does not capture identifiers from its caller's environment nor performs metaprogramming actions like token-pasting or stringification~\cite{gnu-c-preprocessor-stringification}, as C variables lack these features.
\texttt{QUIT} satisfies these premises, because it expands to the literal \texttt{`q'}.
Premises~\ref{pr:var:addr} and~\ref{pr:var:size} verify the macro is not used with the address-of (\texttt{\&}) or \texttt{sizeof} operators, because differences between macro and C scoping rules mean that variables in macro expansions and function bodies have different memory addresses and sizes.

%
%
Premise~\ref{pr:var:not-ice} validates that \texttt{QUIT} is not used where a compile-time constant is required (e.g., in a case label), because C variables are not compile-time constants\footnote{C23, finalized Oct. 2024, permits compile-time constant variables with \texttt{constexpr}~\cite{c23-standard}.}.
%
%
\texttt{QUIT} satisfies Premise~\ref{pr:var:not-ice} because it expands in an \texttt{if} condition (Line~\ref{ln:quit-call}) where compile-time constants are not required.
Premise~\ref{pr:var:const-expr} ensures that \texttt{QUIT}'s definition is a constant expression, since it will be translated to a global variable, per Premise~\ref{pr:var:global}, and C requires global initializers to be constant.
\texttt{QUIT} satisfies Premise~\ref{pr:var:const-expr} because it expands to the character literal \texttt{`q'}.
%

Finally, \tool{} checks Premises~\ref{pr:var:gamma}{-}\ref{pr:var:expr-type} to
ensure that the macro's type is not \texttt{void}---valid only for functions, not variables---and that the macro is \emph{monomorphic}, i.e., all its expansions have the same type, since C variables are monomorphic.
Expanding only to a character type (\texttt{char}), \texttt{QUIT} satisfies these premises.
%
Since \texttt{QUIT} satisfies all premises, \tool{} can follow Rule~\subref{rule:var}'s conclusion to correctly rewrite the macro definition on Line~\ref{ln:quit-def} to a global C variable definition.

\noindent
\begin{minipage}{0.95\linewidth}
\begin{lstlisting}[style=style-c-numbered,escapechar=^]
^\highlightcodegreen{static const char QUIT = 'q';}^
void check_quit(void) {
    if (QUIT == getch()) exit(0);
}
\end{lstlisting}
\end{minipage}

Note that \emph{no other parts of the C source code change}.
\tool{}'s specification ensures that identical syntax can be used at macro invocation sites and the rewritten C variables, functions, or enums, e.g., Line~\ref{ln:quit-call}'s usage of \texttt{QUIT}.

Rule~\subref{rule:enum} identifies macro-to-\texttt{enum} translations.
Unlike Rule~\subref{rule:var}, it identifies invocations used where only compile-time constant expressions are permitted (Premise~\ref{pr:enum:ice}).

\subsection{Translating Function-like Macros}

Rules~\subref{rule:nvf} and~\subref{rule:vf} translate function-like macros to functions that do and do not return a value (\texttt{void} functions), respectively.
Returning value affects function call site semantics.
To demonstrate Rule~\subref{rule:nvf}, consider the function-like macro \texttt{MUL()}, which expands to the expression \texttt{((n)) * ((m))}:

\begin{minipage}{0.9\linewidth}
		\begin{lstlisting}[style=style-c-numbered, caption={\texttt{MUL()} macro
    definition and invocation.},label={lst:mul}]
#define MUL(X, Y) ((X)*(Y))|\label{ln:nvf:mul-def}|
int *new_matrix(int n, int m) {
    return calloc(MUL(n, m), sizeof(int));|\label{ln:nvf:mul-call}|
}
\end{lstlisting}
\end{minipage}

Premises~\ref{pr:nvf:global}{-}\ref{pr:nvf:not-ice} are the same as
those for object-like macros
(Premises~\ref{pr:var:global}{-}\ref{pr:var:not-ice}) explained above in
Section~\ref{sec:macro-to-var}.

Premise~\ref{pr:nvf:expr} verifies that \texttt{MUL()} is an expression, because C function calls and their return values must be expressions.
%
%
\texttt{MUL()} satisfies Premise~\ref{pr:nvf:expr} because it expands to an arithmetic expression.
Premises~\ref{pr:nvf:gamma}{-}\ref{pr:nvf:expr-type} check that \texttt{MUL()} has a monomorphic type, since C functions are monomorphic.
\texttt{MUL()} satisfies premises~\ref{pr:nvf:gamma}{-}\ref{pr:nvf:expr-type} since its one invocation expands to a single integer type.


Premise~\ref{pr:nvf:args} uses Rules~\subref{rule:arg}{-}\subref{rule:arg-list} to translate untyped macro arguments to typed C function arguments.
%
%
Rules~\subref{rule:empty-args} and~\subref{rule:arg-list} translate lists of arguments, and Rule~\subref{rule:arg} translates individual arguments.
For instance, \texttt{MUL()}'s first argument, \texttt{X}.
Rule~\subref{rule:arg} first uses Premise~\ref{pr:arg:alpha} to collect \texttt{X}'s expansions.
%
Premise~\ref{pr:arg:env} ensures that \texttt{X}, like C function arguments, does not capture identifiers from its caller's environment.
%
\texttt{X} satisfies Premise~\ref{pr:arg:env} because it only refers to
\texttt{n}, which is passed as \texttt{MUL()}'s \texttt{X} argument
(Listing~\ref{lst:mul}, Line~\ref{ln:nvf:mul-call}).
Like C variables, C function arguments lack preprocessor metaprogramming capabilities, and C function arguments behave differently from macro arguments when used with address-of (\texttt{\&}) or size-of operators (Section~\ref{sec:macro-to-var}).  Premises~\ref{pr:arg:addr}{-}\ref{pr:arg:not-ice} specify these conditions.



Premise~\ref{pr:arg:cbn} validates that the argument does not use call-by-name
semantics, as function arguments are call-by-value.
Call-by-name semantics evaluate an argument and its side-effects each time it is used in the macro's body, which may be many, or not at all if the macro appears in the unevaluated branch of a short-circuiting operation such as logical and (\texttt{\&\&}).
In contrast, call-by-value semantics evaluate functions' arguments exactly once at their callsites.
\texttt{X} satisfies Premise~\ref{pr:arg:cbn} because the first argument to the
call to \texttt{MUL()} on Line~\ref{ln:nvf:mul-call} of Listing~\ref{lst:mul}
does not have side-effects, and \texttt{MUL()}'s definition does not use
\texttt{X} in a short-circuiting operation.

Finally, Premise~\ref{pr:arg:expr} ensures \texttt{X} is monomorphic, which
\texttt{X} is because it expands to the integer expression \texttt{n}.
Satisfying Rule~\subref{rule:arg}'s premises, both macro arguments \texttt{X}
and \texttt{Y} can be translated following the rewrite rule to \texttt{int X}
and \texttt{int Y}, respectively.
%
Since \texttt{MUL()} and its arguments satisfy Rule~\subref{rule:nvf}, the macro can be translated to:

\begin{lstlisting}[style=style-c-numbered,escapechar=^]
^\highlightcode{static inline int MUL(int X, int Y) \{ return ((X)*(Y)); \}}^
int *new_matrix(int n, int m) {
    return calloc(MUL(n, m), sizeof(int));
}
\end{lstlisting}

As with object-like macros, this translation is a drop-in replacement for the macro and requires no changes to the macro's callsites.
\tool{} also makes \texttt{MUL()}'s translation \texttt{inline} to inline its callsites at compile-time, preserving the original macro's performance.

\subsection{Derivation and Correctness of Rules}
%

\paragraph{Derivation}
The rules in Figure~\ref{fig:translation-rules} were derived from studying
prior work on translating C preprocessor usage to C code.
We studied translations that were correct but limited (mennie) in scope, and
translations that were more complete, but unreliable (vs code, c2rust).
By studying what correct translations checked for, and what incorrect
translations failed to consider, we arrived at the set of overlapping features
between macros and C code shown in Figure~\ref{fig:venn}.
On top of this, we also checked the C preprocessor's ISO specification and user
manual to obtain a complete view of all the features macros can exhibit.
These overlapping features were then used to create the translation rule
premises in Figure~\ref{fig:translation-rules}.

\paragraph{Correctness}
\tool{} accounts for all possible macros when performing its translation,
leaving none unconsidered, and only translates macros satisfying all the
premises of one of Figure~2's rules.
To check these premises for all macros, \tool{} uses the semantic Boolean
properties returned by the macro analyzer Maki~\cite{maki}, which has a test
suite of 95 tests verifying its accuracy~\cite{maki-github}.
Maki's semantic properties cover all macros, ensuring that macros unsupported
by Figure~2's rules are left untranslated, while only those matching a rule's
premises are converted to a C construct.
\tool{}'s translation rules guarantee that at most a single translation will
apply to a macro, preventing the inadvertent translation of unsupported macros,
with all the \tool{} translations analyzed in Section~\ref{sec:evaluation}
being correct.

%% file: figures/functions.tex
\begin{table}[htbp]
\centering
\begin{tabular}{@{}ll@{}}
\toprule
\textbf{Name}                           & \textbf{Condition checked}                   \\ \midrule
\textsc{DefinedGlobally}(\textit{name}) & \textit{name} is defined outside a function. \\
\textsc{CalledByName}($\iota$)          & $\iota$ uses call-by-name semantics.         \\
\textsc{CapturesEnv}($\iota$)           & $\iota$ accesses caller's environment.       \\
\textsc{IsConstExpr}($\iota$)           & $\iota$ is a constant C expression.          \\
\textsc{IsExpression}($\iota$)          & $\iota$ is a C expression.                   \\
\textsc{IsICE}($\iota$)                 & $\iota$ is a compile-time constant (ICE).    \\
\textsc{IsStatement}($\iota$)           & $\iota$ is a not an expression.              \\
\textsc{AddressRequired}($\iota$)       & $\iota$ must be an addressable expression.   \\
\textsc{SizeRequired}($\iota$)          & $\iota$ must be an expression with a size.   \\
\textsc{ConstExprRequired}($\iota$)     & $\iota$ must be a compile-time constant.     \\
\textsc{IsMeta}($\iota$)                & $\iota$ manipulates tokens.                  \\ \bottomrule
\end{tabular}
\caption{Informal function definitions for Figure~\ref{fig:translation-rules}. Each function accepts either a macro \textit{name}, or an invocation or argument expansion $\iota$, as input, and outputs a boolean.}
\label{tab:functions}
\end{table}

%% file: figures/venn-diagrams.tex

\newcommand{\vennsubfigwidth}{0.45\linewidth}
\newcommand{\venngraphicswidth}{\linewidth}

\begin{figure}[ht]
    \centering
    \includegraphics[width=\linewidth]{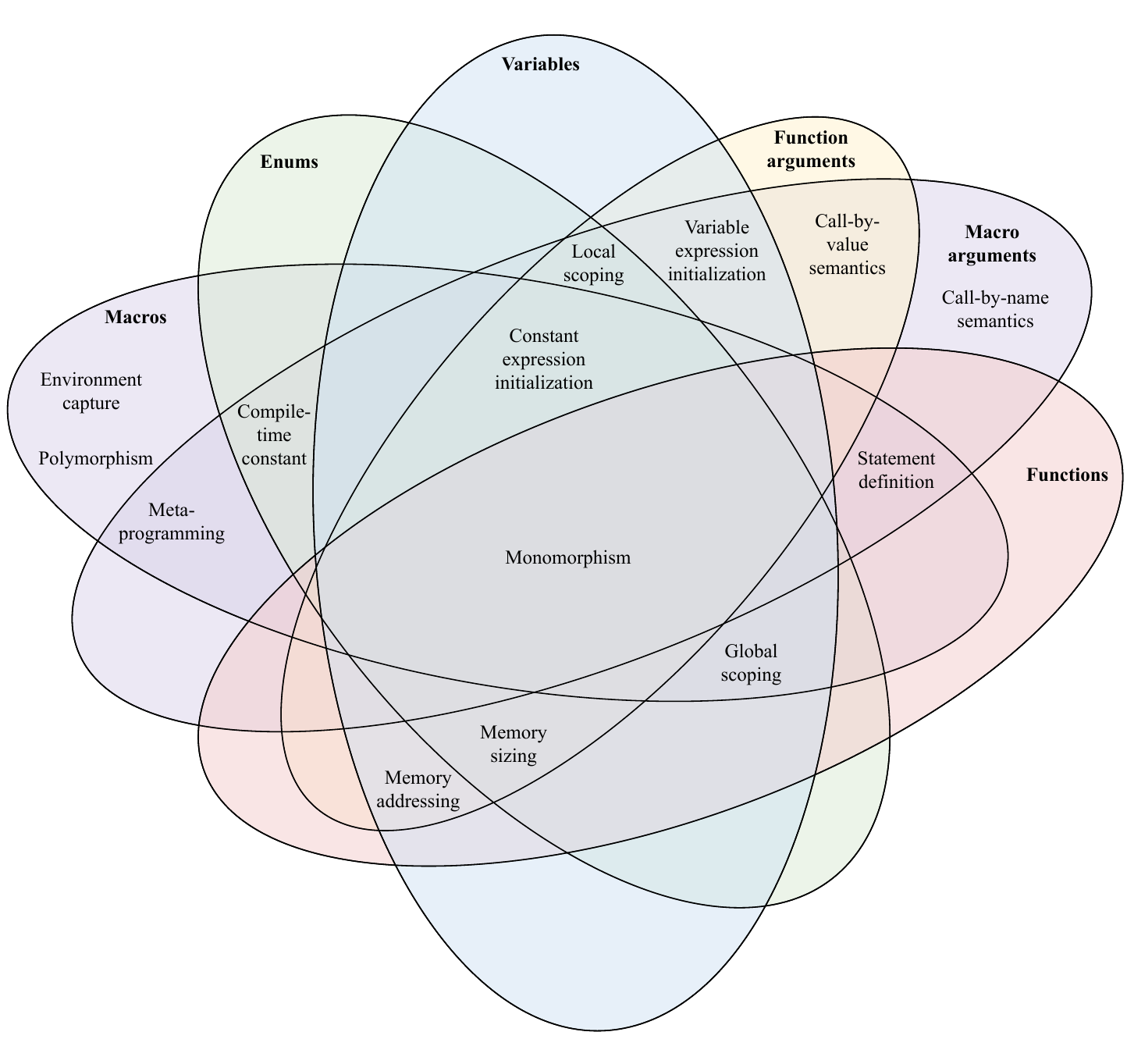}
    \caption{Semantics available to C and macro constructs.}
    \label{fig:venn}
         \Description[Venn diagram comparing semantic features available to C
         and macro constructs.]{Six-way Venn diagram comparing the language
         features available to variables, enums, functions, function arguments,
         macros, and macro arguments. A macro which only use language features
         in the overlapping sections of each diagram behaves the same as a C
         construct, and is therefore safe to translate to that construct by
         replacing its definition with one of the matching construct, without
         needing to alter the macro's callsites as well.}
\end{figure}

%% file: figures/translation-rules.tex
\newcounter{leftpremise}
\newcounter{midpremise}
\newcounter{rightpremise}

\newcommand{\newleftpremise}[1]{\refstepcounter{leftpremise}\label{#1}(\arabic{leftpremise})}
\newcommand{\newmidpremise}[1]{\refstepcounter{midpremise}\label{#1}(\arabic{midpremise})}
\newcommand{\newrightpremise}[1]{\refstepcounter{rightpremise}\label{#1}(\arabic{rightpremise})}

\renewcommand{\theleftpremise}{\arabic{leftpremise}}
\renewcommand{\themidpremise}{\arabic{midpremise}}
\renewcommand{\therightpremise}{\arabic{rightpremise}}

\begingroup{}
    
\renewcommand\thesubfigure{\Roman{subfigure}}
\captionsetup[subfigure]{labelformat=rewriterule, labelsep=colon, singlelinecheck=false}

\begin{figure*}
	\begin{subfigure}[t]{\textwidth}
		\centering{}
		\caption{Object-like macro to global variable}\label{rule:var}
        \begin{tabular}{llllll}
            \newleftpremise{pr:var:global} & $\textsc{DefinedGlobally}(\textit{name})$                           & \setcounter{midpremise}{4}{\newmidpremise{pr:var:addr}} & $\forall \iota \in \mu : \lnot \textsc{AddressRequired}(\iota)$   & \setcounter{rightpremise}{8}{\newrightpremise{pr:var:gamma}} & $\gamma = \Gamma(name)$          \\
            \newleftpremise{pr:var:mu}     & $\mu = M(name)$                                                     & \newmidpremise{pr:var:size}                             & $\forall \iota \in \mu : \lnot \textsc{SizeRequired}(\iota)$      & \newrightpremise{pr:var:gamma-1}                             & $|\gamma| = 1$                   \\
            \newleftpremise{pr:var:env}    & $\forall \iota \in \mu : \lnot \textsc{CapturesEnvironment}(\iota)$ & \newmidpremise{pr:var:not-ice}                          & $\forall \iota \in \mu : \lnot \textsc{ConstExprRequired}(\iota)$ & \newrightpremise{pr:var:type-in-gamma}                       & $type \in \gamma$                \\
            \newleftpremise{pr:var:meta}   & $\forall \iota \in \mu : \lnot \textsc{IsMeta}(\iota)$              & \newmidpremise{pr:var:const-expr}                       & $\forall \iota \in \mu : \textsc{IsConstExpr}(\iota)$             & \newrightpremise{pr:var:expr-type}                           & $type \not\in \{\textbf{void}\}$ \\
			\midrule
			\multicolumn{6}{c}{M, $\Gamma \vdash$ \textbf{\#}\textbf{define} \textit{name} \textit{body} $\leadsto$ \textbf{static const} \textit{type} \textit{name} \textbf{=} \textit{body} \textbf{;}}
		\end{tabular}
	\end{subfigure}\setcounter{leftpremise}{\value{rightpremise}}

	\begin{subfigure}[t]{\textwidth}
		\centering
		\caption{Object-like macro to enum}\label{rule:enum}
        \begin{tabular}{llllll}
            \newleftpremise{pr:enum:global} & $\textsc{DefinedGlobally}(\textit{name})$                           & \setcounter{midpremise}{16}{\newmidpremise{pr:enum:addr}} & $\forall \iota \in \mu : \lnot \textsc{AddressRequired}(\iota)$ & \setcounter{rightpremise}{20}{\newrightpremise{pr:enum:gamma-1}} & $|\gamma| = 1$                                              \\
            \newleftpremise{pr:enum:mu}     & $\mu = M(name)$                                                     & \newmidpremise{pr:enum:size}                              & $\forall \iota \in \mu : \lnot \textsc{SizeRequired}(\iota)$    & \newrightpremise{pr:enum:type-in-gamma}                          & $type \in \gamma$                                           \\
            \newleftpremise{pr:enum:env}    & $\forall \iota \in \mu : \lnot \textsc{CapturesEnvironment}(\iota)$ & \newmidpremise{pr:enum:ice}                               & $\forall \iota \in \mu : \textsc{IsICE}(\iota)$                 & \newrightpremise{pr:enum:type-size}                              & \textsc{SizeOf}($type$) $\le$ \textsc{SizeOf}(\textbf{int}) \\
            \newleftpremise{pr:enum:meta}   & $\forall \iota \in \mu : \lnot \textsc{IsMeta}(\iota)$              & \newmidpremise{pr:enum:gamma}                             & $\gamma = \Gamma(name)$                                         &                                                                  &                                                             \\
			\midrule
			\multicolumn{6}{c}{M, $\Gamma$ $\vdash$ \textbf{\#}\textbf{define} \textit{name} \textit{body} $\leadsto$ \textbf{enum \{} \textit{name} \textbf{=} \textit{body} \textbf{\};}}
		\end{tabular}
	\end{subfigure}\setcounter{leftpremise}{\value{rightpremise}}
    
	\begin{subfigure}[t]{\textwidth}
        \setcounter{subfigure}{2}
		\centering
		\caption{Function-like macro to non-\texttt{void} function}\label{rule:nvf}
		\begin{tabular}{llllll}
			\newleftpremise{pr:nvf:global} & $\textsc{DefinedGlobally}(\textit{name})$                           & \setcounter{midpremise}{28}{\newmidpremise{pr:nvf:size}} & $\forall \iota \in \mu : \lnot \textsc{SizeRequired}(\iota)$      & \setcounter{rightpremise}{33}{\newrightpremise{pr:nvf:type-in-gamma}} & $type \in \gamma$                                     \\
            \newleftpremise{pr:nvf:mu}     & $\mu = M(name)$                                                     & \newmidpremise{pr:nvf:not-ice}                           & $\forall \iota \in \mu : \lnot \textsc{ConstExprRequired}(\iota)$ & \newrightpremise{pr:nvf:expr-type}                                    & $type \not\in \{\textbf{void}\}$                      \\
            \newleftpremise{pr:nvf:env}    & $\forall \iota \in \mu : \lnot \textsc{CapturesEnvironment}(\iota)$ & \newmidpremise{pr:nvf:expr}                              & $\forall \iota \in \mu : \textsc{IsExpression}(\iota)$            & \newrightpremise{pr:nvf:args}                                         & \textit{name}, A, $\Delta \vdash args \leadsto args'$ \\
            \newleftpremise{pr:nvf:meta}   & $\forall \iota \in \mu : \lnot \textsc{IsMeta}(\iota)$              & \newmidpremise{pr:nvf:gamma}                             & $\gamma = \Gamma(name)$                                           &                                                                       &                                                       \\
            \newleftpremise{pr:nvf:addr}   & $\forall \iota \in \mu : \lnot \textsc{AddressRequired}(\iota)$     & \newmidpremise{pr:nvf:gamma-1}                           & $|\gamma| = 1$                                                    &                                                                       &                                                       \\
			\midrule
			\multicolumn{6}{c}{M, $\Gamma$, A, $\Delta \vdash$ \textbf{\#}\textbf{define} \textit{name} \textbf{(} \textit{args} \textbf{)} \textit{body} $\leadsto$ \textbf{static}~\textit{type}~\textit{name}~\textbf{(} \textit{args'} \textbf{)} \textbf{\{}~\textbf{return}~\textit{body}~\textbf{;}~\textbf{\}}}
		\end{tabular}
	\end{subfigure}\setcounter{leftpremise}{\value{rightpremise}}

	\begin{subfigure}[t]{\textwidth}
		\centering
		\caption{Function-like macro to \texttt{void} function}\label{rule:vf}
		\begin{tabular}{llllll}
\newleftpremise{pr:vf:global} & $\textsc{DefinedGlobally}(\textit{name})$                           & \setcounter{midpremise}{41}{\newmidpremise{pr:vf:size}} & $\forall \iota \in \mu : \lnot \textsc{SizeRequired}(\iota)$                            & \setcounter{rightpremise}{46}{\newrightpremise{pr:vf:type-in-gamma}} & $type \in \gamma$                                     \\
\newleftpremise{pr:vf:mu}     & $\mu = M(name)$                                                     & \newmidpremise{pr:vf:not-ice}                           & $\forall \iota \in \mu : \lnot \textsc{ConstExprRequired}(\iota)$                       & \newrightpremise{pr:vf:expr-type}                                     & $type \in \{\textbf{void}\}$                          \\
\newleftpremise{pr:vf:env}    & $\forall \iota \in \mu : \lnot \textsc{CapturesEnvironment}(\iota)$ & \newmidpremise{pr:vf:expr-stmt}                         & $\forall \iota \in \mu : \textsc{IsExpression}(\iota) \lor \textsc{IsStatement}(\iota)$ & \newrightpremise{pr:vf:args}                                          & \textit{name}, A, $\Delta \vdash args \leadsto args'$ \\
\newleftpremise{pr:vf:meta}   & $\forall \iota \in \mu : \lnot \textsc{IsMeta}(\iota)$              & \newmidpremise{pr:vf:gamma}                             & $\gamma = \Gamma(name)$                                                                 &                                                                       &                                                       \\
\newleftpremise{pr:vf:addr}   & $\forall \iota \in \mu : \lnot \textsc{AddressRequired}(\iota)$     & \newmidpremise{pr:vf:gamma-1}                           & $|\gamma| = 1$                                                                          &                                                                       &                                                       \\
			\midrule
			\multicolumn{6}{c}{M, $\Gamma$, A, $\Delta \vdash$ \textbf{\#}\textbf{define} \textit{name} \textbf{(} \textit{args} \textbf{)} \textit{body} $\leadsto$ \textbf{static}~\textbf{void}~\textit{name}~\textbf{(} \textit{args'} \textbf{)} \textbf{\{}~\textit{body}~\textbf{;}~\textbf{\}}}
		\end{tabular}
	\end{subfigure}\setcounter{leftpremise}{\value{rightpremise}}

	\begin{subfigure}[t]{0.30\textwidth}
		\centering
		\caption{Empty macro argument list}\label{rule:empty-args}
		\begin{tabular}{c}
			\midrule
			\textit{name}, A, $\Delta \vdash \qquad{} \leadsto \qquad{} $
		\end{tabular}
	\end{subfigure}
	\begin{subfigure}[t]{0.45\textwidth}
		\centering
		\caption{Non-empty macro argument list}\label{rule:arg-list}
		\begin{tabular}{llll}
			\newleftpremise{pr:args:first} & \textit{name}, A, $\Delta \vdash arg \leadsto arg'$   &
			\newleftpremise{pr:args:rest}  & \textit{name}, A, $\Delta \vdash args \leadsto args'$ \\
			\midrule
			\multicolumn{4}{c}{\textit{name}, A, $\Delta \vdash arg\textbf{,}~args \leadsto arg'\textbf{,}~args'$}
		\end{tabular}
	\end{subfigure}

	\begin{subfigure}[t]{\textwidth}
		\centering
		\caption{Non-empty macro argument}\label{rule:arg}
		\begin{tabular}{llllll}
            \newleftpremise{pr:arg:alpha} & $\alpha = \textrm{A}(name, arg)$                                       & \setcounter{midpremise}{55}\newmidpremise{pr:arg:size} & $\forall \iota \in \alpha : \lnot \textsc{SizeRequired}(\iota)$      & \setcounter{rightpremise}{59}\newrightpremise{pr:arg:delta} & $\delta = \Delta(name, arg)$     \\
            \newleftpremise{pr:arg:env}   & $\forall \iota \in \alpha : \lnot \textsc{CapturesEnvironment}(\iota)$ & \newmidpremise{pr:arg:not-ice}                         & $\forall \iota \in \alpha : \lnot \textsc{ConstExprRequired}(\iota)$ & \newrightpremise{pr:arg:delta-1}                            & $|\delta| = 1$                   \\
            \newleftpremise{pr:arg:meta}  & $\forall \iota \in \alpha : \lnot \textsc{IsMeta}(\iota)$              & \newmidpremise{pr:arg:cbn}                             & $\forall \iota \in \alpha : \lnot \textsc{CalledByName}(\iota)$      & \newrightpremise{pr:arg:type-in-delta}                      & $type \in \delta$                \\
            \newleftpremise{pr:arg:addr}  & $\forall \iota \in \alpha : \lnot \textsc{AddressRequired}(\iota)$     & \newmidpremise{pr:arg:expr}                            & $\forall \iota \in \mu : \textsc{IsExpression}(\iota)$               & \newrightpremise{pr:arg:expr-type}                          & $type \not\in \{\textbf{void}\}$ \\
			\midrule
			\multicolumn{6}{c}{\textit{name}, A, $\Delta \vdash arg \leadsto type~arg$}
		\end{tabular}
	\end{subfigure}

	\caption{Formal rules for \tool{}'s translation of macros to C code.}\label{fig:translation-rules}
    \Description[Formal rules for translating macros and their arguments to C code.]{A set of formal rules for translating C Preprocessor macros to C variables, enums, functions, and C macro arguments to function arguments. Each rule consists of a set of premises that must be true for the rule to apply, a set of input variables that are necessary to check if the rule's premises are satisfied, and a conclusion specifying the translation to apply when all the rule's premises are true.}
\end{figure*}

\endgroup{}

%% file: implementation.tex
\begin{figure}
	\begin{center}
		\includegraphics[width=0.95\linewidth]{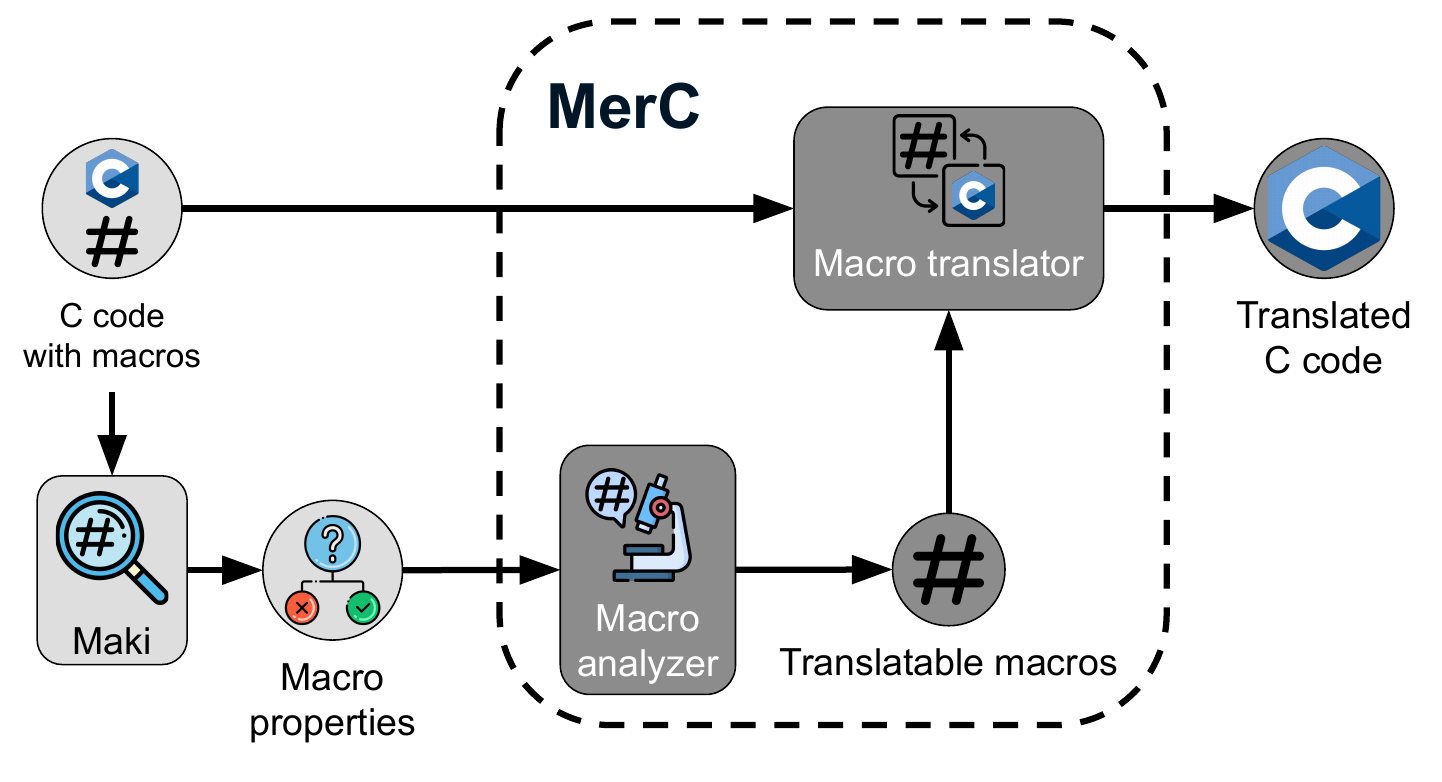}
	\end{center}

	\caption{
		\tool{} architecture diagram.
		Circles are data.  Dark gray rectangles are novel contributions.  Maki is prior work.
	}\label{fig:architecture-diagram}
    
    \Description[\tool{}' architecture.]{A diagram describing \tool{}'s implementation architecture. First a user passes a C source file to a pre-existing macro analyzer, Maki, to obtain the semantic properties of its macros. Then they pass the original source file along with its semantic macro properties to \tool{} for it to determine which macros to translate and how. Finally, \tool{} replaces the definitions of translatable macros with the appropriate C constructs, and writes the resulting code back to the original C file. Untranslatable macros and all macro callsites are left unchanged.}

\end{figure}

\noindent{}Figure~\ref{fig:architecture-diagram} shows the architecture of
\tool{}, which inputs unpreprocessed C source code and emits unpreprocessed C
source with supported macros translated to C constructs.
\tool{} is implemented in 952 source lines of Python code.
%


\tool{} works by first using the macro analyzer Maki~\cite{maki} to compute,
for all AST-aligned macro invocations, a set of fine-grained Boolean properties
about the macro's semantics.
The artifact~\footref{fn:artifact} file \texttt{macros.py} then maps these
Boolean properties to the functions in Table~\ref{tab:functions}.
These helper functions are used by
\texttt{predicates/inter\-face\_equivalent.py} to implement
Figure~\ref{fig:translation-rules}'s rule premises for choosing which
translation rule, if any, to apply to a given macro.
Finally, the file \texttt{macrotranslator.py} implements each translation
rule's conclusion to conduct the actual translation of each macro.

%
Because \tool{}'s translation rules preserve the same syntax of macro
callsites, \tool{} only ever needs to modify the definitions of supported
macros to translate them to C constructs.
Moreover, \tool{} retains macros' performance benefits by translating macros to
\texttt{inline} functions.
Downstream translators can then convert these inline functions to compile-time
functions in the target language, such as Rust's own macros or
\texttt{\#[inline]} functions.

%% file: benchmark.tex
We introduce \benchmark{}, the first benchmark specifically focused on macro translation.
\benchmark is designed to be realistic, varied, focused, and usable.
We constructed \benchmark{} by randomly sampling macro usage from widely-varying, real-world C programs to ensure representativeness, independence from any particular translation techniques, and statistical significance.


\subsection{Design Principles}\label{sec:benchmark:principles}
We adhered to the following principles when creating \benchmark{}.

\paragraph{Realism}
\benchmark{}'s test cases should reflect real-world macro usage to best represent a translator's performance on C code.
%
%
For realism we base \benchmark{} on the 26 real-world programs studied in the seminal work on macro analysis~\cite{ernst-et-al}, removing 8 programs that use C++ or whose dependencies are no longer available\footnote{gcc, gnuchess, gs, mosaic, plan, rasmol, workman, and zephyr.}.
We compensate for the removed programs by adding 5 modern ones: Linux, Lua, Python, SQLite, and FFMpeg, for a total of \numbenchmarkprograms{} programs, increasing the benchmark to 100,711 macro definitions\footnote{See the file \texttt{programs.txt} in our public artifact for a listing of each program's name, version, source lines of code, and number of invoked and sampled macro definitions.}.
%

%
We sample macros with stratified random sampling, where the strata are the source programs.
We calculate the number of macros sampled from each program based on its number of definitions, so that all programs contribute at least one macro to \benchmark{}.
Otherwise, small programs would likely contribute no macros, and large programs like the Linux kernel would dominate.
To ensure representativeness, we choose a statistically-significant sample size of \numtestcases{}, for a 95\% confidence level and 5\% margin of error.

Macros in real-world code do not appear in isolation, but are surrounded by declarations, statements, and expressions.
\benchmark{} preserves this context by including the declarations of any variables, functions, types, or other symbols the sampled macro references, along with expressions or statements invoking the sampled macro.


\paragraph{Variety}
\benchmark{}'s test cases should contain diverse macro usage to evaluate a translator's ability to handle the many kinds of macros real-world C uses.
To achieve variety, we sample macros from a selection of C software spanning diverse software categories, including databases, programming languages, and window management.
Furthermore, by stratifying macros by their source programs we ensure \benchmark{} includes macros from all programs, even small ones, since simple random sampling would likely exclude macros from small programs altogether.
%
After sampling, we verified by hand that \benchmark{} contains myriad types of macros that have been previously documented by researchers~\cite{mennie-and-clarke}, including named constants, function name aliases, generic functions, and metaprogramming macros. 22\% are function-like and 78\% are object-like, reflecting the distribution found by prior research~\cite{ernst-et-al}.
Additionally, we check for diverse semantics~\cite{ernst-et-al,demacrofication,maki}, such as variable capture, token-pasting, control-flow modifications, and more.
The complete listing of macros, originating programs, and additional metadata can be found in our accompanying artifact\footref{fn:artifact}.


\paragraph{Focus}
\benchmark{} is designed to specifically measure macro translation capacity, without penalizing tools for lack of whole program translation or build system analysis. 
%
In particular, LLMs' limited context windows and cloud service input token limits preclude prompting them with entire programs, which may comprise millions of lines of code, only to evaluate macro translation capacity.
%
Similarly, C programs use build automation to set compiler flags and use header files to share common macro definitions across C files.
%
To evaluate macro translation with whole program translation would require a tool to handle these pragmatic issues in order to even observe macro usage.
%
In contrast, \benchmark{} enables a baseline comparison of macro translation in isolation from whole program translation pragmatics.
To accomplish this, we hand-slice the macro definition and invocations into a single C file per macro, preserving any local language usage that surrounds the macro and is used within the macro to retain its semantic context.




\paragraph{Usability}
\benchmark{} is made to be readily accessible and usable for other researchers to evaluate macro translation to C or other languages.
%
%
%
We organize \benchmark{}'s test cases into directories based on the
programs we sampled them from and include build automation that researchers and tool designers can use to integrate and automate their tool's translation of the benchmark.
%
We make each test case a single file so that AI
researchers may directly prompt models to perform translations without also
providing header files, or whole programs as inputs.
Finally, we include the evaluated tools' translations (\Cref{sec:evaluation}) along with the semi-automatically validated results of the evaluation, including which translations are correct or incorrect, failure type, locations, descriptions, and other metadata, so that future researchers developing macro translations have a baseline for comparison.
\benchmark{} is available publicly and is included in the accompanying artifact\footref{fn:artifact} for review.

\subsection{Benchmark Creation Procedure}
We used the following procedure to convert each sampled macro into a \benchmark{} test case.
For each test case we created one C source file, and copied into it a single sampled macro's definition and invocation sites.
Most macros were only invoked a handful of times, and for these test cases we included all invocations.
But a small number (12\%) of macros were invoked thousands of times across many C files, and to include them all would produce a file too large for input to LLMs.
So for these macros, we used a random number generator to select a random sample of the macros invocations, and copied these into the test case file.
Each such case is described in the macro benchmark listing in the accompanying artifact\footref{fn:artifact}.
%
%
For macros invoked within C functions or other constructs, to maintain
realistic semantics we copied the macro's surrounding language constructs,
along with any types, variables, additional macros, and other symbols that the
benchmark macro relies on.
%
We included context recursively; if a macro depends on the declaration of some
type A that itself depends on type B, we included declarations of both A and B
in the benchmark.
Similarly, since macros may be called within other macros, we preserve the
calling context of nested benchmark macros by partially expanding all
macro invocations until reaching the sampled benchmark macro, so that the
benchmark macro's invocation is explicit.


%% file: evaluation.tex
\hypersetup{linkcolor=black}
We investigate these six research questions about \tool{} and LLMs.
\begin{enumerate}[label={\textbf{RQ\arabic{enumi}}},
		leftmargin=\widthof{\textbf{RQ\arabic{enumi}}}+\labelsep{}]

	\item \hyperref[rq:attempt-rate]{How many translations do tools attempt?}

    \item \hyperref[rq:failure-rate]{How many translations are failures?}

    \item \hyperref[rq:time]{How fast are translations?}

    \item \hyperref[rq:tag-team]{How fast and effective are \tool{}-LLM Tag Teams?}

	\item \hyperref[rq:failures]{What are LLMs' translations failures?}

	\item \hyperref[rq:prompting-strategy]{How does prompting strategy affect LLM translations?}


\end{enumerate}

\subsection{LLM Selection}

We chose the following LLMs: \gpt{}~\cite{gpt-4o} because it is
popular and likely to be used by developers in practice;
\claude{}~\cite{claude-3.5-sonnet} because it is designed for solving coding
tasks~\cite{claude-3.5-sonnet-excels-at-coding}; and
\oonepreview{}~\cite{o1-preview} for its focus on reasoning~\cite{o1-reasoning}.
%
We use the GitHub Copilot Visual Studio Code extension~\cite{copilot-vscode} to
translate macros with \gpt{} and \claude{}, because Copilot is a popular tool
that reflects developer LLM usage.
\gpt{} is its current default model.
%
Copilot recently added support for \claude{}, a model that reportedly
``excels at coding tasks''~\cite{claude-3.5-sonnet-excels-at-coding}.  Since macro translation is a coding task, we also use Copilot with \claude{}.
%
%
%
Finally, we include \oonepreview\footnote{The full o1 release occurred after we started experiments.\label{fn:oone-availability}} to represent LLMs with reportedly superior reasoning~\cite{o1-reasoning}.
Copilot currently imposes strict daily usage limits on this model~\cite{copilot-o1-usage-limits}, so we use OpenAI's own API~\cite{openai-api-docs} to evaluate it.

\subsection{Experimental Setup and Methods}
\label{sec:setup}

We performed the LLM translations using cloud services (Copilot and OpenAI's) and performed \tool{} translations on an off-the-shelf laptop with an
8-core Intel i7-11857G processor and 16GB of RAM.

We performed LLM translations following these steps: (1) start a new chat with
the model, (2) prompt the model to translate a test case, (3) record the
translation time, and (4) save the translation in a new file.
All LLMs were given the same zero-shot~\cite{zero-shot-definition} prompt:


\begin{center}
\begingroup
\addtolength\leftmargini{-0.75cm}
\begin{quote}
	translate this code to pure C, translating the macro \textit{MACRO\_NAME}
	to an enum, function, or variable without changing program semantics or
	callsites. You are NOT ALLOWED to change the callsites of the macro or
	create any new macros. Therefore, if it is not possible to translate the
	macro without preserving the callsites and semantics, do not perform any
	translation. Report a confidence level in percentage in the accuracy of the
	translation in a comment at the end within the code.


	\textit{CODE}

\end{quote}
\endgroup
\end{center}

\noindent{}%
\texttt{MACRO\_NAME} and \texttt{CODE} are replaced with the name of the macro
to translate and the code (wrapped in backticks) containing the macro from
\benchmark{}, respectively.
We developed the prompt iteratively, asking LLMs to translate Maki's~\cite{maki}
macro analysis unit tests and picking the one that worked best.
We chose zero-shot prompting because it is a popular prompting
strategy~\cite{schulhoff2025promptreportsystematicsurvey} that developers are
likely to use in practice due to its ease-of-use and
flexibility~\cite{practical-survey-on-zero-shot-design}.
We determined the elapsed translation time for \gpt{} and \claude{} from
Copilot's debugging output, subtracting the time the model received the request
from the time it finished responding to it.
For \oonepreview, we instrumented the request script to collect wall-clock time.
To translate test cases with \tool{}, we instrumented \benchmark{}'s build
system to run \tool{} on each source file and record the time taken on each test
case.

LLM translations were run once per test case.
Temperature settings used for \gpt{} and \claude{} were the default provided by
the Copilot VS Code extension since Copilot provided no way to change that
setting at the time our evaluation was
conducted~\cite{vscode-temperature-setting}.
To remain fair and consistent with using the default temperature setting for
\gpt{} and \claude{}, we used the default temperature setting for \oonepreview{}
as well.

We began checking the correctness of all translations by automatically checking
for compile-time correctness with GCC.
Most translation failures, 80\%, were compile-time errors found using automated checking with GCC.
The remaining 20\% of failures not suffering from compile-time errors were
checked (\hyperref[rq:failure-rate]{RQ2}) and labeled
(\hyperref[rq:failures]{RQ5}) by a team of students trained by a co-author.
The same co-author then validated all results, so that each translation was
checked at least twice.
When a labeling discrepancy arose between student and co-author opinion, the
co-author had the final say; though the number of such discrepancies was small.
All our analyses can be found in the accompanying artifact\footref{fn:artifact}.


\begin{figure*}
    \begin{subfigure}[t]{0.31\textwidth}
        \centering
        \includegraphics[width=\linewidth]{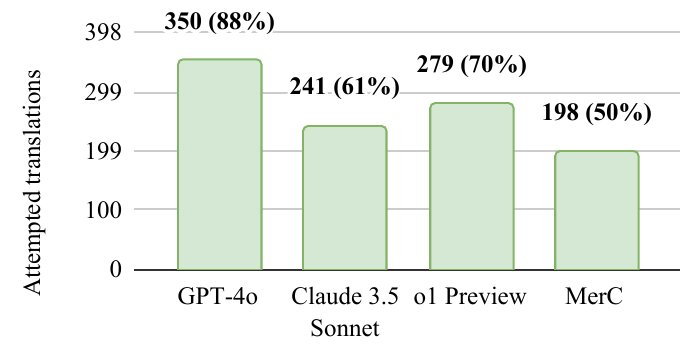}
        \caption{Attempted translations.}\label{fig:attempted}
        \Description[%
        Number of test cases that each tool attempted to translate, out of 398 test cases.
        ]%
        {
        Number of translations that each tool attempts, out of 398 test cases.
        \gpt{} translates the most test cases, 350, for an attempt rate of 88\%.
        \claude{} translates 241 test cases for an attempt rate of 61\%.
        \oonepreview{} translates 279 test cases for an attempt rate of 70\%.
        \tool{} translates the fewest test cases, 198,, for an attempt rate of 50\%.
        }
    \end{subfigure}%
    \quad
    \begin{subfigure}[t]{0.31\textwidth}
        \centering
        \includegraphics[width=\linewidth]{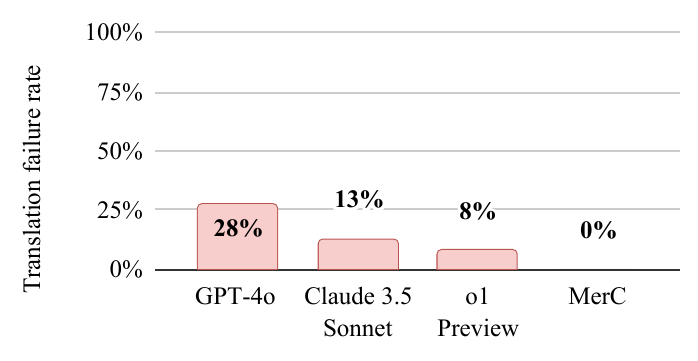}
        \caption{Translation failure rates.}\label{fig:failed}
        \Description[%
        Percentage of each tool's attempted translations that were failures.
        ]%
        {
        Percentage of each tool's attempted translations that were failures.
        28\% of \gpt{}'s attempted translations were failures, the greatest failure rate of any tool evaluated.
        13\% of \claude{}'s attempted translations were failures.
        8\% of \oonepreview{}'s attempted translations were failures.
        \tool{} had no failures.
        }
    \end{subfigure}%
    \quad
    \begin{subfigure}[t]{0.31\textwidth}
        \centering
        \includegraphics[width=\linewidth]{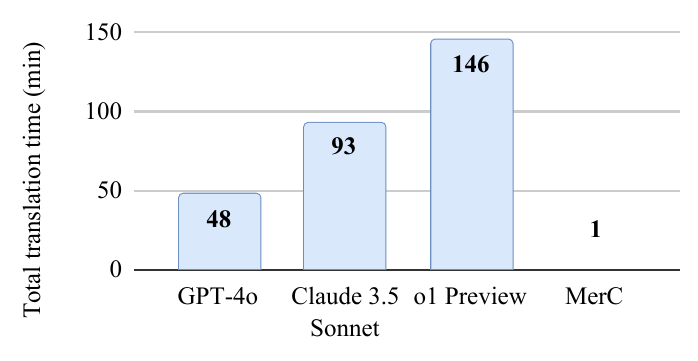}
        \caption{Total translation times (minutes).}
        \label{fig:times}
        \Description[%
        Total translation times for each tool.%
        ]%
        {\gpt{} performed all its translations in 48 minutes.
        \claude{} performed all its translations in 93 minutes.
        \oonepreview{} performed all its translations in 146 minutes.
        \tool{} performed all its translations in 37 seconds.
        }
    \end{subfigure}
    \caption{Attempt rates, failure rates, and translation times for LLMs and \tool{} when translating \benchmark{}'s test cases.}
\end{figure*}

\subsection{RQ1: How Many Translations do Tools Attempt?}%
\label{rq:attempt-rate}

The attempt rate is how many \benchmark{} test cases a tool attempted to
translate, regardless of correctness.
We consider a translation attempted when the tool translates the macro
definition into a C declaration (variable, function, etc.).
%
%
For instance, an LLM may refuse to translate or leave the input
unchanged.

Figure~\ref{fig:attempted} is the total number and percent of attempted translations for each tool.
GPT-4o has the greatest attempt rate of 88\%, while MerC has the lowest attempt rate, 50\%.
%
Claude 3.5 Sonnet and o1 Preview, designed for better coding and reasoning
tasks~\cite{claude-3.5-sonnet-excels-at-coding,o1-reasoning}, have attempt
rates between that of GPT-4o and MerC, attempting 61\% and 70\% of test
cases, respectively.
%
%
While all LLMs emit more translations than \tool{}, their translations are not always correct.

\calloutbox{}{%
    \textbf{RQ1:} All LLMs attempt more translations than \tool{}, but their correctness is not guaranteed.
}

\subsection{RQ2: How Many Translations are Failures?}
\label{rq:failure-rate}

A failure is a translation that produces invalid or semantically different C
code.  The failure rate is the percentage of attempted translations that fail.
A perfectly accurate translation would attempt all translations (100\% attempt
rate) without failures (0\% failure rate).


\Cref{fig:failed} shows failure rates for each tool on \benchmark{}.
\tool{} attempts the fewest translations, 198, but emits no incorrect translations, for a failure rate of 0\%.
In contrast, all LLMs emit many wrong translations, between 8\% and 28\%.
\gpt{} in particular has the highest failure rate of 28\%, with 97 of its 350
translations being incorrect.
The reasoning-style LLMs (\claude{} and \oonepreview{}) have lower failure rates than \gpt{} (13\% and 8\%, respectively), but also attempt fewer translations than \gpt{} as well (241 and 279 translations, respectively).

The downside of any non-zero failure rate, no matter how small, is that developers need to validate all LLM translations, \emph{even the correct ones}, to ensure the correctness of the results, because LLMs do not distinguish correct from incorrect translations in their output.
Although we prompted LLMs to report their confidence in the translation, they
almost always reported a high confidence level, even for incorrect
translations.
Finally, 30 macros from \benchmark{} were never translated by any approach.
These macros use features like
token-pasting~\cite{gnu-c-preprocessor-token-pasting} that lack a direct C
equivalent.

\calloutbox{}{%
    \textbf{RQ2:} \tool{} had no failures but translates only those macros for
    which it is specifically designed.  In contrast, all LLMs made more
    translations but always produced some incorrect ones, forcing developers to
    manually verify all translations.
	

}

\subsection{RQ3: How Fast Are Translations?}%
\label{rq:time}

\Cref{fig:times} presents the total time each tool spent translating all test cases.
\Cref{fig:times} shows that, despite running on a personal laptop, \tool{} is orders of magnitude faster than all LLMs, performing all translations in less than a minute.
In contrast, \gpt{}  takes 48 minutes to perform all translations, while the more accurate model, \claude{}, takes an hour and a half, and \oonepreview{}, the most accurate model, spends the most time at two and a half hours.

\calloutbox{}{%
    \textbf{RQ3:} The rule-based \tool{} is orders of magnitude faster than LLMs (seconds vs. hours).
	%
}

\subsection{RQ4: How Fast and Effective are \tool{}-LLM Tag Teams?}%
\label{rq:tag-team}

\begin{figure}
    \centering
    \includegraphics[width=.9\linewidth]{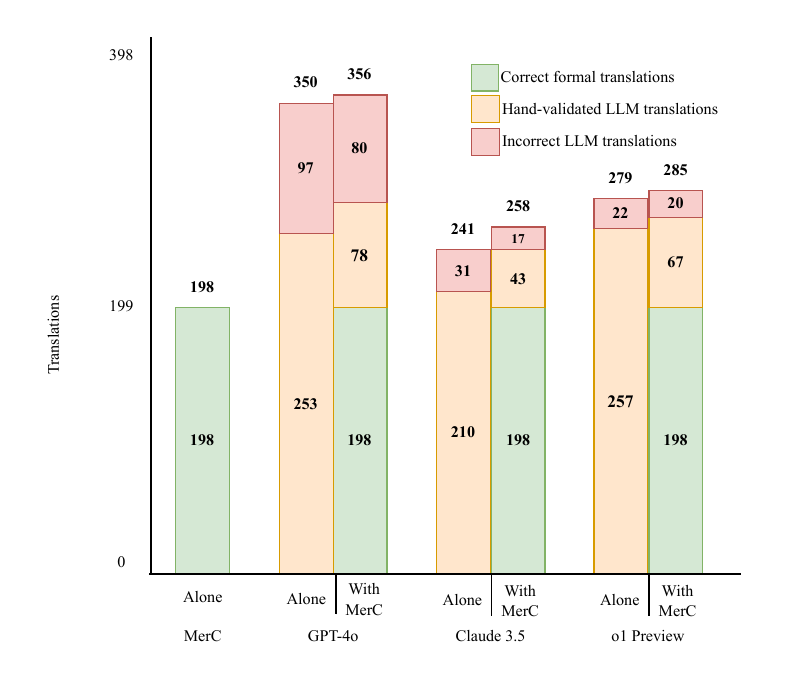}
    \caption{Types of translations performed by \tool{}, LLMs, and \tool{}-LLM tag teams.}%
    \label{fig:combined}
    \Description[Types of translations performed by \tool{}, LLMs, and \tool{}-LLM tag teams.]{\tool{} made 198 correct translations requiring no validation.
    When run alone, \gpt{} made 253 correct translations requiring hand-validation, and 97 incorrect translations, for a total of 350 translations.
    The \tool{}-\gpt{} tag team made 198 correct translations requiring no validation, 78 translations requiring validation, and 80 incorrect translations, for a total of 356 translations.
    When run alone, \claude{} made 210 correct translations requiring hand-validation, and 31 incorrect translations, for a total of 241 translations.
    The \tool{}-\claude{} the tag team made 198 correct translations requiring no validation, 43 translations requiring validation, and 17 incorrect translations, for a total of 258 translations.
    When run alone, \oonepreview{} made 257 correct translations requiring hand-validation, and 22 incorrect translations, for a total of 279 translations.
    The \tool{}-\oonepreview{} tag team made 198 correct translations requiring no validation, 67 translations requiring validation, and 20 incorrect translations, for a total of 285 translations.}
\end{figure}

%
Neither \tool{} or LLMs are better in both number of translations and
failure rate.
But they have complementary strengths: \tool{} is fast and perfectly accurate
while LLMs make more translations.
%
We unite these complementary strengths into a \emph{translation tag team} that
first translates with \tool{} then ``tags in'' an LLM for the remaining macros
\tool{} does not support.
The tag team approach reaps greater benefits than either technique alone,
decreasing translation time while increasing both the number and accuracy of
translations.

The tag team works better than \tool{} or LLMs alone, because LLMs incorrectly
translate some macros that fall within \tool{}'s rules for common macro cases
(there are 28 macros that \tool{} gets right that at least one LLM tool gets
wrong).
By avoiding LLM translation of macros that \tool{} already supports, developers
need validate fewer translations for correctness.
And by using LLMs to translate the remaining macros that \tool{} does not
support, the tag team approach translates more macros overall.

\Cref{fig:combined} compares \tool{} and the LLMs' attempts and failures to
those of \tool{}-LLM tag teams.
%
%
Because \tool{} only produces correct translations, the tag team
approach only has an average \hybridaveragefailurerate{} failure rate (7\% to
22\%), a \hybridaveragelowerfailureratethanllms{} reduction in failures
compared to using LLMs alone.
Moreover, since the tag team approach supplements \tool{}'s 198 translations
with extra LLM translations, the total number of attempted translations is on
average \hybridaveragegreatertranslationsthantool{} higher (30\% to 80\%) than
\tool{}'s 198, with an average of
\hybridaveragegreatercorrecttranslationsthantool{} more correct translations to
boot.
Finally, by translating 198 macros correctly with \tool{} first before using LLMs to translate the remaining \benchmark{} macros, the number of LLM translations needing hand-validation drops by 66\% on average for tag teams compared to using LLMs alone. 

\begin{figure}
    \centering
    \includegraphics[width=.8\linewidth]{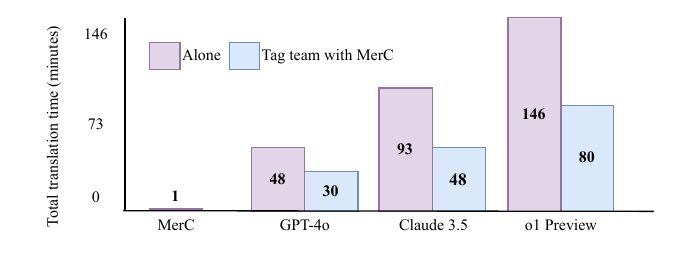}
    \caption{Total translation times for \tool{}, LLMs, and tag teams.  Lower is better. \tool{}'s 37 seconds rounds to 1 minute.}%
    \label{fig:times-combined}
    \Description[Total translation times for \tool{}, LLMs, and tag teams. Lower is better.]{\tool{} translates all test cases in 37 seconds, which rounds to 1 minute.
    When run alone, \gpt{} translates all test cases in 48 minutes.
    The \tool{}-\gpt{} tag team translates all test cases in 30 minutes.
    When run alone, \claude{} translates all test cases in 93 minutes.
    The \tool{}-\claude{} tag team translates all test cases in 48 minutes.
    When run alone, \oonepreview{} translates all test cases in 146 minutes.
    The \tool{}-\oonepreview{} tag team translates all test cases in 80 minutes.}
\end{figure}

Similarly, \tool{} is orders of magnitude faster than LLMs, translating 50\% of the benchmark in under a minute.
Therefore, restricting LLM use to a subset of macros greatly reduces overall run-time.
%
\Cref{fig:times-combined} shows the total time each LLM spent translating the test cases not first translated by \tool{}.
%
The \tool{}-LLM tag teams run in nearly half the time as the LLMs in isolation, with LLMs alone taking 48 minutes to 2.5 hours, and the \tool{}-LLM tag teams taking only 30 minutes to 1.3 hours.

We also characterize the gap between what macros \tool{} correctly translates
and the macros that LLMs correctly translate. There were 103 macros (69
object-like, 34 function-like) correctly translated by LLMs, but not attempted
by MerC. MerC did not attempt these macros because it failed to find an
abstraction-preserving translation. To determine why, we took a random sample of
20 of these 103 macros (10 object-like, 10 function-like).

5 of the 20 sampled macros referenced a symbol declared after the macro's
definition, meaning that in order to translate the macro, a developer would need
to rearrange the order of definitions in their code, which MerC considers as an
abstraction-breaking translation since it requires more work than simply
replacing a macro's definition with its translated definition.
4 of the 20 sampled macros were invoked in a short-circuiting expression (e.g.,
\texttt{||} or \texttt{\&\&}), precluding MerC translation because in the general
case macros invoked in such expressions cannot be translated to function calls
without possibly breaking program semantics due to differences between function
call-by-value and macro call-by-name calling conventions.
3 of the 20 sampled macros were not translated because of limitations with MerC's
implementation: these macros made nested calls to other macros, which is a
feature that MerC's underlying macro analyses engine, Maki, does not fully
support.
2 of the 20 sampled macros were object-like macros defined to a construct other than
a constant C expression, preventing MerC from translating them in an
abstraction-preserving way to variables or enums, but still allowing for LLMs to
translate these macros to functions.
The remaining macros were not translated, because they were metaprogramming
macros, were used in a preprocessor conditional, or expanded to a C declaration
rather than a C value.

\calloutbox{}{%
    \textbf{RQ4:}
    Using both \tool{} and LLMs averages \hybridaveragelowerfailureratethanllms{} fewer failures than LLMs alone and \hybridaveragegreatertranslationsthantool{} more translations than \tool{} alone.
    %
}

\subsection{RQ5: What Are LLMs' Translation Failures?}%
\label{rq:failures}


We semi-automatically checked the correctness of each LLM and \tool{} translation along the following criteria, recording the first error found.

\paragraph{Compile-time behavior} Translated code must compile successfully
without syntax or type errors.
We automatically test for compilation errors with \texttt{gcc}, marking
translations that fail to compile as compile-time failures.
\newcommand{\macroname}{\texttt{FRAME\_VISIBLE\_P()}}
For example, consider \oonepreview{}'s translation of the Emacs macro
\macroname{}, which fails to compile because the translated version of
\macroname{} references the struct \texttt{frame} before it is declared:

\begin{minipage}{0.38\linewidth}
\begin{lstlisting}[style=style-c-no-numbered, caption={\texttt{FRAME\_VISIBLE\_P()}.}, xleftmargin=0cm, label=lst:compile-time-failure-original, escapechar=^, aboveskip=1em, frame=single]
^\highlightcode{\#define FRAME\_VISIBLE\_P(f) \\}^
    ^\highlightcode{(f)->visible}^

struct frame {
    unsigned visible : 2;
};
\end{lstlisting}
\end{minipage}
\quad
\begin{minipage}{0.50\linewidth}
\begin{lstlisting}[style=style-c-no-numbered, caption={Failed \oonepreview{} translation.}, xleftmargin=0cm, label=lst:compile-time-failure-translation, escapechar=^, aboveskip=1em, frame=single]
^\highlightcode{static inline unsigned}^
^\highlightcode{FRAME\_VISIBLE\_P(struct frame *f) \{}^
    ^\highlightcode{return f->visible;}^
^\highlightcode{\}}^
struct frame {
    unsigned visible : 2;
};
\end{lstlisting}
\end{minipage}

%
The translated code fails to compile because the function translation of
\macroname{} references the struct type \texttt{frame}, which is not declared
until later in the program.
Macros may be defined before any types they reference because macros are
dynamically scoped, but C functions, being statically scoped, must be defined
after any types they reference.
%

\paragraph{Runtime behavior} All the translated macro's invocations must
produce the same result as the original test case before translation.
We compared each macro to its translated counterpart, and marked changes in
macro behavior (e.g., changes to side-effects) as failures.
\renewcommand{\macroname}{\texttt{Vafter\_delete}}
For instance, \gpt{}'s translation of the Emacs macro \macroname{} changes the
macro's runtime behavior by turning the macro into a completely new variable,
whereas the original macro aliased an existing one (identifiers shortened for
space):

\begin{minipage}{0.44\linewidth}
\begin{lstlisting}[style=style-c-no-numbered, caption={Original \macroname{} macro.}, xleftmargin=0cm, label=lst:runtime-failure-original, escapechar=^, aboveskip=1em, frame=single]
^\highlightcode{\#define Vafter\_delete \textbackslash }^
  ^\highlightcode{globals.f\_Vafter\_delete}^
struct emacs_globals {
  Lisp_Object f_Vafter_delete;
};
extern struct emacs_globals
  globals;
\end{lstlisting}
\end{minipage}
\quad
\begin{minipage}{0.44\linewidth}
\begin{lstlisting}[style=style-c-no-numbered, caption={Failed \gpt{} translation.}, xleftmargin=0cm, label=lst:runtime-failure-translation, escapechar=^, aboveskip=1em, frame=single]
struct emacs_globals {
  Lisp_Object f_Vafter_delete;
};

extern struct emacs_globals
  globals;
^\highlightcode{Lisp\_Object Vafter\_delete;}^
\end{lstlisting}
\end{minipage}

The translation fails to preserve the original program's behavior because the
translation of \macroname{} no longer aliases
\texttt{globals.f\_Vafter\_delete}, but is rather an entirely new variable.
%

\paragraph{Non-functional behavior} The translated macro must preserve the
program's memory layout and preprocessor semantics.  For instance, translations
that alter \texttt{\#ifdef} behavior or mutate unions to structs fail to
preserve non-functional program behavior, although these only represent a
minority of failures (\nonfunctionalfailurespercent{} of all LLM failures).
%
\renewcommand{\macroname}{\texttt{CONFIG\_MOV\_MUXER}}
For example, \claude{}'s translation of the FFmpeg macro \macroname{} converts a
macro used in a preprocessor conditional into a C enum, changing the
conditional's output:

\begin{minipage}{0.43\linewidth}
\begin{lstlisting}[style=style-c-no-numbered, caption={Original \macroname{} macro.}, xleftmargin=0cm, label=lst:non-functional-failure-original, escapechar=^, aboveskip=1em, frame=single]
^\highlightcode{\#define CONFIG\_MOV\_MUXER 1}^
#if CONFIG_MOV_MUXER ^//^ true
const FFOutputFormat 
ff_mov_muxer = {
    .p.name = "mov"
};
#endif
\end{lstlisting}
\end{minipage}
\quad
\begin{minipage}{0.45\linewidth}
\begin{lstlisting}[style=style-c-no-numbered, caption={Failed Claude translation.}, xleftmargin=0cm, label=lst:non-functional-failure-translation, escapechar=^, aboveskip=1em, frame=single]
^\highlightcode{enum \{ CONFIG\_MOV\_MUXER = 1 \};}^
#if CONFIG_MOV_MUXER ^//^ false
const FFOutputFormat
ff_mov_muxer = {
    .p.name = "mov"
};
#endif
\end{lstlisting}
\end{minipage}

The translated code  alters the original program's non-functional
behavior because it moves \macroname{} from the preprocessor namespace into the
C namespace, altering the output of the program's \texttt{\#if}.
Since the original program defines \macroname{} to 1, the conditional will
evaluate to true and declare the variable \texttt{ff\_mov\_muxer}.
But because the translated code converts the macro to an enum, which is not a
preprocessor symbol, the \texttt{\#if} will evaluate to false, and
\texttt{ff\_mov\_muxer} will never be declared.
%



\begin{table}[ht]
    \centering
    \begin{tabular}{lccccc}
        \toprule
                       & \tool{} & \gpt{} & Claude 3.5 & \oonepreview{} \\
        \midrule
        Compile Time   & 0 & 76 & 26 & 18 \\
        Runtime        & 0 & 4  & 2  & 2  \\
        Non-functional & 0 & 17 & 3  & 2  \\
        \midrule
        Total          & 0 & 97 & 31 & 22 \\
        \bottomrule
    \end{tabular}
    \caption{Types of failed translations.}%
    \label{tab:failure-reasons}
\end{table}


\Cref{tab:failure-reasons} presents the number of test cases that each tool
translated incorrectly.
\tool{} had no failures.
For LLMs, the majority of translation failures were compile-time errors.
Roughly half of all compile-time errors for \gpt{} and \claude{} were due to
the use of non-constant values where constant values were required.
This reflects a misunderstanding of C semantics and the common use of
macros to define constants: macros get expanded before C compilation, so they
can be used where C variables cannot.

For \oonepreview{}, however, the majority of compile-time errors (14 out of 18)
were due to type errors, specifically undeclared references.
On one hand, this indicates that \oonepreview{} is better at avoiding syntactic
errors, but on the other it also suggests that the model misunderstands how the
preprocessor interacts with C's typing rules.
For example, Listing~\ref{lst:compile-time-failure-translation} presents an
\oonepreview{}-translated macro argument which suffers a type error because it
references a type declared later in the program.
MerC avoids compile-time errors because its translation rules encode both C
syntax and semantics.
\calloutbox{}{%
        %
	\textbf{RQ5:} \tool{} correctly translates all attempted test cases, while
	LLMs' failed translations typically have compile-time errors.
	%
}

\subsection{RQ6: How Does Prompting Strategy Affect LLM Translations?}%
\label{rq:prompting-strategy}

\begin{figure}
    \centering
    \includegraphics[width=0.8\linewidth]{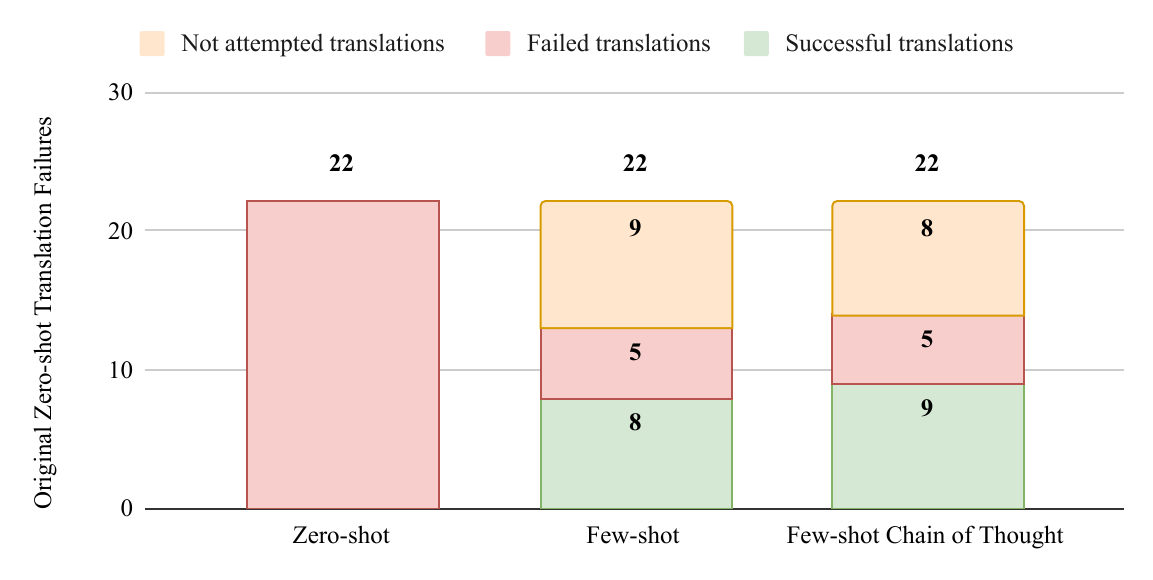}
    \caption{Comparison of the number of \oonepreview{}'s failed translations under various prompting strategies.
    }%
    \label{fig:o1-prompting}
    
    \Description[Comparison of the number of \oonepreview{}'s failed
    translations under zero-shot, few-shot, and few-shot chain-of-thought
    prompting strategies.]{Comparison of the number of \oonepreview{}'s failed
    translations under zero-shot, few-shot, and few-shot chain-of-thought
    prompting strategies. With zero-shot prompting, \oonepreview{} emits 22
    incorrect translations. With few-shot prompting, 8 of the 22 originally
    failed translations become successes, and 9 of them become non-attempted
    translations. With few-shot chain-of-thought prompting, 9 of these failed
    translations become successes, and 8 become non-attempted translations.}
\end{figure}


All LLM translations were conducted using a zero-shot prompt (\Cref{sec:setup}).
%
%
%
%
To evaluate whether a different prompting strategy would improve on failed LLM translations, we evaluated two additional popular prompting strategies, few-shot~\cite{few-shot-learning} and few-shot chain-of-thought (CoT)~\cite{neurips2022_9d560961}.
Few-shot prompts include example solutions of the task to
solve~\cite{few-shot-definition}, while few-shot CoT also
adds explanations of the reasoning steps followed in the examples~\cite{cot-definition}.
We evaluate prompt engineering on \oonepreview{} specifically.
%
Since it showed the best accuracy (\hyperref[rq:failure-rate]{RQ2}) and has a reportedly high reasoning capacity~\cite{o1-reasoning}, it stands to benefit the most from prompt  engineering~\cite{o1-prompt-engineering}.
The few-shot and CoT prompt templates can be found in our accompanying artifact\footref{fn:artifact} in \prompttemplatepath{}. 

%

Figure~\ref{fig:o1-prompting} compares the number of \oonepreview{} translation
failures with a zero-shot prompting strategy to the number of translation
failures under few-shot and few-shot CoT prompting strategies.
\oonepreview{} made 22 incorrect (out of 279 attempted) test case translations
with the zero-shot prompt (\Cref{rq:failure-rate}).
With few-shot, \oonepreview{} correctly translated eight of these cases, still incorrectly translated five of them, and did not attempt the remaining nine.
Few-shot CoT was similar, only correctly translating one additional case that was not attempted with the few-shot prompt. 
%
While the prompting strategies helped reduce failed translations they also made \oonepreview{} more selective in avoiding translation altogether.


%

\calloutbox{}{%
    \textbf{RQ6:} Few-shot and few-shot CoT reduced failed translations but also
    resulted in fewer attempts on prior failed translations.
}

%% file: threats.tex
\subsection{Internal Threats}

%
%


\paragraph{\tool{} translation scope}
\tool{} only translates macros expanding to complete C constructs, and not those expanding to C syntax fragments. 
%
%
However, users can translate nested macro invocations by repeatedly running \tool{} on its own output to ``peel back'' layers of nested macro usage until no more translations occur.
%
%

\paragraph{LLM selection}
Section~\ref{sec:evaluation}'s results on how well LLMs translate macros should represent real-world developer LLM usage.
To this end, we evaluate \gpt{} because it is popular, \claude{} because it is designed for coding tasks~\cite{claude-3.5-sonnet-excels-at-coding}, and \oonepreview{}~\cite{o1-reasoning} for its focus on reasoning.
To further reflect developer LLM usage, we evaluate \gpt{} and \claude{} with the GitHub Copilot Visual Studio Code Extension~\cite{copilot-vscode}.
%
Although Copilot also supports \oonepreview{}, it currently imposes strict daily usage limits on the model~\cite{copilot-o1-usage-limits}, so we use OpenAI’s own API~\cite{openai-api-docs} to evaluate it\footref{fn:oone-availability}.

\paragraph{LLM prompting}
%
%
To ensure Section~\ref{sec:evaluation} evaluates LLMs with a quality prompt, we designed our prompt iteratively by asking LLMs to translate small random samples of macros from Maki's~\cite{maki} macro analysis test cases until all LLMs performed the task as instructed.

\paragraph{LLM performance results}
To ensure Section~\ref{rq:time} accurately portrays LLM performance on macro translation, we ran the LLMs using cloud services, since real developers are likely to lack dedicated AI machines, and instead use LLMs directly from their providers.
%

\paragraph{Independence of the benchmark}
%
%
%
%
To prevent biasing \benchmark{} towards \tool{}, we designed the benchmark independently from \tool{}'s translation rules, with randomly selected test cases to avoid even inadvertently specializing \tool{} to \benchmark{}.
\benchmark{} was created after \tool{} was developed, and by a separate author.
%

\paragraph{Inter-rater agreement}
%
%
We did not use a formal inter-rater agreement metric when identifying and classifying failed translations in Section~\ref{sec:evaluation}.
To address this threat, a student and a co-author checked each translation, and then met to resolve discrepancies and arrive at a final mutually agreed label for each translation.
Afterwards, the same co-author verified that all translation labels were consistent.

\subsection{External Threats}
%
%

\paragraph{Benchmark representativeness}
To ensure Section~\ref{sec:evaluation}'s findings about macro translation generalize to real developer scenarios, we draw \benchmark{}'s test cases from the \numbenchmarkprograms{} programs described in Section~\ref{sec:benchmark}, which vary in size, complexity, and application domains.

\paragraph{Translation of macros from compile-time to run-time abstractions}
%
By translating macros to C, \tool{} gives downstream translators the ability to
choose how to translate each macro in a way that best preserves its
abstraction.
But translating macros into functions threatens to reduce program performance,
since macros are compile-time abstractions, and functions are run-time
abstractions.
\tool{} mitigates this threat by converting all C macros into inline functions,
which are expanded at their callsites by the compiler to remove the overhead
associated with entering and exiting function calls.
Downstream translation tools can then convert these inline functions into
compile-time constructs in the target language to not only preserve the
original macro's abstraction, but also its performance.
For example, in order to preserve runtime performance, c2rust could convert
\tool{}-translated macros into inline Rust functions~\cite{rust-inline}, or
even Rust macros~\cite{rust-macros}.
\paragraph{Translating whole programs}
%
%
%
\benchmark{} represents real-world macro translation tasks, but factors out practical challenges of whole program translation so as to enable evaluating LLMs, whose limited input windows do not support large C programs as translation inputs.
%
%
%
In contrast, although \tool{} is designed to translate single compilation units and not whole programs, it can be applied to whole programs by translating each of their compilation units individually.
%
We use this strategy to explore how well \tool{} translates macros appearing organically in the \numbenchmarkprograms{} real-world C programs from which \benchmark was drawn.
Translations were run in parallel on a server with 2x AMD EPYC 7742 64-Core and 512GB RAM.
\tool{} took 9.5 days to analyze all programs, most of which was spent on the Linux kernel, which required 8.5 days.
The median analysis time was about 2 minutes per program.
After analysis, \tool{} took 2 minutes to convert all \mercattemptedrealworld{} macros, with a median translation time of 1 second.
%
We check correctness by compiling the translated programs and running their built-in test suites.
All \numbenchmarkprograms{} successfully compile and pass tests after \tool{} translation.
%
Out of \totalrealworldmacros{} total invocations, \tool{} finds and translates \mercattemptedrealworld{} of them (\mercattemptraterealworld{}).
This percentage is low partly because some macro definitions are defined in header files shared across compilation units, and so can't be translated without additional specifications for the semantics of whole program macro usage.
Another reason is that \tool{} only translates macros that are not called within other macros; this limitation can be overcome by repeatedly using \tool{} to translate programs until no more translations occur.
This approach would always terminate, as macros have bounded recursion~\cite{gnu-c-preprocessor-self-referential-macros}.

%% file: related-work.tex
\paragraph{Translating C preprocessor macros}
Mennie and Clarke described the first automated macro translation tool in
2004~\cite{mennie-and-clarke}.
Their tool is not publicly available, and only translates constant object-like
macros into C variables and enums, whereas \tool{} also translates function-like
macros into C functions.
Macroscope~\cite{astec} translates C preprocessor code to ASTEC code, but fails
to help translate macro-laden C code to safer languages since ASTEC is itself a
preprocessing language.
Kumar et al.\ discussed ``rejuvenating'' C++ programs by translating C
preprocessor macros to C++~\cite{demacrofication}.
Unlike \tool{}, their translation is not fully automatic, translates to C++ instead of C, and is sometimes incorrect because it ignores macro invocation semantics.
%
%
%
%
%
%
C2rust can translate some C preprocessor macros to
Rust~\cite{c2rust-issue-304}, but this feature has correctness
issues~\cite{c2rust-issue-803, c2rust-issue-829}.
Visual Studio can refactor certain macro definitions into constant expression declarations~\cite{vs-2017-automatic-macro-refactoring}, but this feature has bugs~\cite{vs-2017-automatic-macro-refactoring-unaddressed-bug}. 
Macroni~\cite{macroni-github,macroni-blog-post} translates C preprocessor macros to LLVM MLIR~\cite{mlir}, but not to C.
%
%

\paragraph{Translating C preprocessor conditionals}
%
Tools exist for translating C preprocessor conditionals~\cite{creconfigurator,
hercules, sugarc, configuration-lifting, variability-encoding}, but they do not
translate macros, and therefore suffer the same issues as other C translators.
Such tools expand macros in C code before translating it, preventing downstream C translators from preserving macros in their translations.
%
%
\tool{} could be used to translate macros first before translating preprocessor conditionals to enable macro-aware variability analysis and verification.

\paragraph{Analyzing C preprocessor usage}
Badros and Notkin presented the first preprocessor-aware C source code analyzer, Pcp\textsuperscript{3}~\cite{pcp3}, in 2000.
%
%
Ernst et al.\ used PCp\textsuperscript{3} to conduct the first empirical study of C preprocessor usage~\cite{ernst-et-al}.
Although they found 75\% of macros to be constants or expressions, this provides
an incomplete picture of macro translatability since their analysis only
considered macro definitions, and ignored invocation semantics.
%
Dietrich presented CppSig for inferring macros' function signatures
from their invocations~\cite{cppsig}, and inferred return types and
argument types for over 50\% of Linux kernel macro definitions.
%
%
Pappas and Gazzillo constructed a framework of 26 Boolean properties
for measuring macro portability, and built on CppSig to implement it as a
tool called Maki~\cite{maki}.
\tool{} in turn uses Maki's properties as input for deciding if and how to
translate macros to C language constructs.

\paragraph{Translating code with LLMs}
There exist many code-generating LLMs~\cite{starcoder, polycoder, santacoder,
codegen, codet5, codex, claude-3.5-sonnet, o1-preview}, with some models
designed specifically for translation~\cite{codegeex}.
Pan et al. compared the effectiveness of seven models for translating code in
seven datasets spanning four programming languages~\cite{lost-in-translation},
and found \gpt{} to emit the greatest percentage of correct translations.
\Cref{rq:failure-rate} shows that the newer \claude{} and \oonepreview{} models
boast even lower translation failure rates than \gpt{}, demonstrating how these
more recent models improve on prior ones.



%% file: acknowledgments.tex
This work is supported in part by the National Science Foundation under grant
CCF-1941816.
We thank Adam Betinsky, Kenneth Valladares, and Ehigie Ekata for helping check
and label the macro translations in \Cref{sec:evaluation}.
We thank the anonymous reviewers for their feedback, which helped improve the
paper's quality.

%% file: data-availability.tex

We make \tool{}, \benchmark{}, and our experimental scripts and results publicly
available in our anonymized artifact~\cite{artifact}.
The artifact contains all macros used to create \benchmark{}, their originating
programs, how many of their invocations are included in \benchmark{}, and
additional metadata.
This metadata includes each tool's translation speed, whether it translated each
macro correctly, reasons why failed macro translations were unsuccessful, and
what C constructs each tool attempted to translate each macro to.
The artifact also provides our zero-shot, few-shot and chain-of-thought prompt
templates in \prompttemplatepath{}.
%